\documentclass[twocolumn]{autart}    

\usepackage{graphicx}          
\usepackage{graphics} 
\usepackage{epsfig} 
\usepackage{mathptmx} 
\usepackage{times} 
\usepackage{amsmath} 
\usepackage{amssymb}  
\usepackage{floatrow}
\usepackage{enumerate}
\usepackage{mathrsfs}
\usepackage{amsfonts}

\DeclareMathOperator{\Tr}{Tr}

\begin{document}

\begin{frontmatter}

\title{Fault Tolerant Quantum Filtering and Fault Detection for Quantum Systems\thanksref{footnoteinfo}} 

\thanks[footnoteinfo]{The preliminary version of this paper has been partly presented at the 34th Chinese Control Conference and SICE Annual Conference, 28-30 July 2015, Hangzhou, China. This work was supported by the Australian Research Council (DP130101658, FL110100020). Corresponding author D.~Dong. Tel. +61-2-62686285.
Fax +61-2-62688443.}

\author[ADFA]{Qing Gao}\ead{qing.gao.chance@gmail.com},
\author[ADFA]{Daoyi Dong}\ead{daoyidong@gmail.com},    
\author[ADFA]{Ian R. Petersen}\ead{i.r.petersen@gmail.com}               

\address[ADFA]{School of Engineering and Information Technology, University of New South Wales, Canberra, ACT, 2600, Australia}  


\begin{abstract}                          
This paper aims to determine the fault tolerant quantum filter and fault detection equation for a class of open quantum systems coupled to a laser field that is subject to stochastic faults. In order to analyze this class of open quantum systems, we propose a quantum-classical Bayesian inference method based on the definition of a so-called quantum-classical conditional expectation. It is shown that the proposed Bayesian inference approach provides a convenient tool to simultaneously derive the fault tolerant quantum filter and the fault detection equation for this class of open quantum systems. An example of two-level open quantum systems subject to Poisson-type faults is presented to illustrate the proposed method. These results have the potential to lead to a new fault tolerant control theory for quantum systems.
\end{abstract}

\end{frontmatter}

\section{Introduction}
The theory of filtering, which in a broad sense is a scheme considering the estimation of the system states from noisy signals and/or partial observations, plays a significant role in modern engineering science. A filter propagates our knowledge about the system states given all observations up to the current time and provides optimal estimates of the system states. From the fundamental postulates of quantum mechanics, one is not allowed to make noncommutative observations of quantum systems in a single realization or experiment. Any quantum measurement yields in principle only partial information about the system. This fact makes the theory of quantum filtering extremely useful in measurement based feedback control of quantum systems, especially in the field of quantum optics (\cite{Rouchon2015}, \cite{Wiseman2009}). A system-probe interaction setup in quantum optics is used as the typical physical scenario concerning the extraction of information about the quantum system from continuous measurements (\cite{Belavkin1992},  \cite{Gardiner2000}). The quantum system under consideration, e.g., a cloud of atoms trapped inside a vacuum chamber, is interrogated by probing it with a laser beam. After interaction with the electromagnetic radiation (laser), the free electrons of the atoms are accelerated and can absorb energy. This energy is then emitted into the electromagnetic field as photons which can be continuously detected through a homodyne detector (\cite{Wiseman2009}). Using the continuous integrated photocurrent generated by the homodyne detector one can conveniently estimate the atomic observables. To find the optimal estimates is then precisely the goal of quantum filtering theory. A very early approach to quantum filtering was presented in a series of papers by Belavkin dating back to the early 1980s (\cite{Belavkin1980}, \cite{Belavkin1992}), which was developed in the framework of continuous nondemolition quantum measurement using the operational formalism from Davies's precursor work (\cite{Davies1969}). In the physics community, the theory of quantum filtering was also independently developed in the early 1990s (\cite{Carmichael1993}), named ``quantum trajectory theory'' in the context of quantum optics.

Particular emphasis is given to the work by Bouten \emph{et al.} (2007) where quantum probability theory was used in a rigorous way and a quantum filter for a laser-atom interaction setup in quantum optics was derived using a quantum reference probability method. A basic idea in quantum probability theory is an isomorphic equivalence between a commutative subalgebra of quantum operators on a Hilbert space and a classical (Kolmogorov) probability space through the spectral theorem, from which any probabilistic quantum operation within the commutative subalgebra can be associated with its classical counterpart. The complete quantum probability model is treated as the noncommutative counterpart of Kolmogorov's axiomatic characterisation of classical probability. Similar to the classical case (\cite{Bertsekas2002}), the optimal estimate of any observable is given by its quantum expectation conditioned on the history of continuous nondemolition quantum measurements of the electromagnetic field. The quantum filter was derived in terms of $It\hat o$ stochastic differential equations using a reference probability method.

In practice, classical randomness may be introduced directly into the system dynamics of quantum systems (\cite{Ruschhaupt2012}). For example, the system Hamiltonian of a superconducting quantum system may contain classical randomness due to the existence of stochastic fluctuations in magnetic flux or gate voltages (\cite{Dong2015}). A spin system may be subject to stochastically fluctuating fields that will introduce classical randomness into the system dynamics (\cite{Dong2012}). For an atom system subject to a laser beam, the occurrence of stochastic faults in the laser device may cause the introduction of classical randomness into the dynamics of the atom system (\cite{Viola2003}, \cite{Khodjasteh2005}). For an open quantum system, the system may evolve randomly and the system dynamics may involve two kinds of randomnesses, i.e., \emph{quantum randomness} due to intrinsic quantum indeterminacy and \emph{classical randomness} arising from the imprecise behaviour of macroscopic devices. In order to solve this issue, Bouten \emph{et al.} (2009) presented an approach to analyzing quantum observables containing classical random information. By using quantum spectral theorem, a classical random variable was equivalently represented by a quantum observable in a commutative quantum probability space on an external Hilbert space. As a result, a random observable can be interpreted by compositing an operator-valued function with this quantum observable and can be well defined on an enlarging quantum probability space. In order to estimate classical random parameters from quantum measurements, joint quantum and classical statistics were also considered in literature using the concept of ``hybrid'' classical-quantum density operator, see e.g., (\cite{Dotsenko2009}, \cite{Gambetta2001}, \cite{Kato2013}, \cite{Negretti2013},  \cite{Somaraju2012}, \cite{Tsang2009a}, \cite{Tsang2009b}, \cite{Tsang2010}). In this paper, we concentrate on a class of open quantum systems subject to stochastic faults, aiming at deriving the fault tolerant quantum filtering equation and the fault detection equation. In order to achieve this goal, we consider an approach to uniformly analyzing quantum observables and classical random variables. First, the isomorphic equivalent relationship between a set of random observables equipped with a quantum-classical expectation operation and a classical probability space model is determined. Then a quantum-classical conditional expectation is considered using the associated classical concept, based on which a Bayes formula is obtained. This Bayesian inference method provides a convenient tool to simultaneously derive the fault tolerant quantum filter and fault detection equations for this class of systems.

This paper is organized as follows. Section 2 describes the class of open quantum systems under consideration in this paper. Section 3 is devoted to statistical interpretation of quantum observables containing information of classical random parameters. In Section 4, the fault tolerant quantum filter and fault detection equations are simultaneously derived for open quantum systems using a Bayesian inference method. An example of two-level quantum systems with Poisson-type faults is illustrated. Section 5 concludes this paper.

\section{Heisenberg Dynamics of Open Quantum Systems}
In this work, we concentrate on an open quantum system that has been widely investigated in quantum optics (\cite{Wiseman2009}, \cite{Qi2013},
\cite{Handel2005}). The quantum system under consideration is a cloud of atoms in weak interaction with an external laser probe field which is continuously monitored by a homodyne detector (\cite{Bouten2007},
\cite{Mirrahimi2007}). Such a quantum system can be described by quantum stochastic differential equations driven by quantum noises $B(t)$ and $B^{\dagger}(t)$ (\cite{Wiseman2009}). The dynamics of the quantum system are described by the following quantum stochastic differential equation\footnote[1]{We have assumed $\hbar$=1 by using atomic units in this paper.}:
\begin{eqnarray}
dU(t)&=&\left\{\left(-iH(t)-\frac{1}{2}L^{\dagger}L\right)dt\right. \nonumber\\
&&\hspace{2cm}+LdB^{\dagger}(t)-L^{\dagger}dB(t)\Bigg\}U(t), \label{gquantum2}
\end{eqnarray}
with initial condition $U(0)=I$ and $i=\sqrt{-1}$. Here $U(t)$ describes the Heisenberg-picture evolution of the system operators and $H(t)$ is the system Hamiltonian. In terms of the system states, if $\pi_0$ is a given system state, we write $\rho_0=\pi_0\otimes \left|\upsilon \right>\left<\upsilon \right|$, where $\left|\upsilon \right>$ represents the vacuum state. The system operator $L$, together with the field operator $b(t)=\dot B(t)$ models the interaction between the system and the field. From quantum $It\hat o$ rule, one has (\cite{Gardiner2000})
\begin{eqnarray*}
&&dB(t)dB^{\dagger}(t)=dt,\\
&&dB^{\dagger}(t)dB(t)=dB(t)dB(t)=dB^{\dagger}(t)dB^{\dagger}(t)=0.
\end{eqnarray*}
The atom system and the laser field form a composite system and the Hilbert space for the composite system is given by $\mathcal{H}_{\mathcal{S}}\otimes \mathcal{E}=\mathcal{H}_{\mathcal{S}}\otimes \mathcal{E}_{t]}\otimes \mathcal{E}_{(t}$ where we have exhibited the continuous temporal tensor product decomposition of the Fock space $\mathcal{E}=\mathcal{E}_{t]}\otimes \mathcal{E}_{(t}$ into the past and future components (\cite{Belavkin1992}, \cite{Holevo1991}). It is assumed that $\dim(\mathcal{H}_{\mathcal{S}})=n<\infty$. The atomic observables are described by self-adjoint operators on $\mathcal{H}_{\mathcal{S}}$. Any system observable $X$ at time $t$ is given by $X(t)=j_t(X)=U^{\dagger}(t)(X\otimes I)U(t)$. It is noted that (\ref{gquantum2}) is written in $It\hat o$ form, as will all stochastic differential equations in this paper.

In practice, the system Hamiltonian may change randomly because of, e.g., faulty control Hamiltonians that appear in the system dynamics at random times (\cite{Viola2003},
\cite{Khodjasteh2005}) or random fluctuations of the external electromagnetic field (\cite{Ruschhaupt2012}, \cite{Dong2015}). In this case, the system Hamiltonian can be described by a Hermitian operator $H(F(t))$ that depends on some classical stochastic process $F(t)$. Using the quantum $It \hat o$ rule (\cite{Hudson1984}), one has $d(U^{\dagger}(t)U(t))=d(U(t)U^{\dagger}(t))=0$, which implies that $U(t)$ is a \emph{random unitary operator} and $X(t)=j_t(X)$ is a \emph{random observable}, both depending on the stochastic process $F(t)$. In this paper, for simplicity we still write $U(t)$ instead of the functional form $U(F,t)$. One can conclude that the commutativity of observables is preserved, that is, $[j_t(A), j_t(B)]=0$ if $[A, B]=0$ where $A,B$ are two system observables in $\mathcal{H}_{\mathcal{S}}$. Here the commutator is defined by $[A, B]=AB-BA$. In addition, from (\ref{gquantum2}) one can see that $U(t)$ depends on $B(t')$ and $B^{\dagger}(t')$, $0\leq t'< t$, since the increments $dB(t)$ and $dB^{\dagger}(t)$ point to the future evolution. Consequently,
\begin{equation}
[U(t), dB(t)]=[U(t), dB^{\dagger}(t)]=0. \label{gquantum3}
\end{equation}
Similarly, the time evolution operator $U(t,s)=U(t)U^{\dagger}(s)$ from time $s$ to time $t$ depends only on the field operators $dB(s')$ and $dB^{\dagger}(s')$ with $s\leq s' \leq t$. Thus,
\begin{equation}
[U(t,s), B(\tau)]=[U(t,s), B^{\dagger}(\tau)]=0, \tau \leq s. \label{gquantum4}
\end{equation}
In quantum experiments, generally measurement is performed on the field. Using homodyne detectors, the observation process is given by $Y(t)=j_t(Q(t))=U^{\dagger}(t)(I\otimes Q(t))U(t)$ where $Q(t)=B(t)+B^{\dagger}(t)$ is the real quadrature of the input field. The operator $Q(t)$ commutes with itself at different times, i.e., $[Q(t), Q(s)]=0$. When the field is initialized in the vacuum state, $Q(t)$ is isomorphically equivalent to a real Wiener process (\cite{Gardiner2000}). Combing (\ref{gquantum3}) and (\ref{gquantum4}) with the fact that $[I\otimes Q(t), X\otimes I]=0$, it is easy to show that: (i) $[Y(t), Y(s)]=0$ at all times $s, t$ and (ii) $[Y(s), X(t)]=0, \forall s\leq t$. These two properties guarantee that (i) $Y(t)$ can be continuously monitored, and (ii) it is possible to obtain the conditional statistics of an observable $X(t)$ based on the history of $Y(t)$. In addition, by using the quantum $It\hat o$ rule, one has 
\begin{equation}
dY(t)=U^{\dagger}(t)(L+L^{\dagger})U(t)dt+dQ(t), \label{gquantum5}
\end{equation}
from which $Y(t)$ looks like $j_t(L+L^{\dagger})=U^{\dagger}(t)(L+L^{\dagger})U(t)$ with a noise $Q(t)$.

\section{Statistical Interpretation of Random Observables}
Like the case we have discussed in Section 2, in many applications classical random variables may be introduced into quantum system Hamiltonian and make the system's evolution depend on some classical random variables. In such a case, both quantum and classical randomnesses will be involved in the system dynamics. An approach to analyzing both quantum and classical random variables using quantum probability theory was proposed in \cite{Bouten2009} to compute the filter equation in the presence of random feedback control signal. In this paper, we consider the fault tolerant quantum filtering problem for a class of open quantum systems subject to classical stochastic faults. In order to solve this problem, we consider a way of uniformly analyzing quantum and classical random variables using a Bayes inference method for calculating joint quantum-classical statistics. This method provides a convenient tool to solve the fault tolerant quantum filtering problem that is the focus of this paper. In this section, we provide a brief introduction to quantum probability theory and present a brief analysis on quantum-classical Bayes inference, which is used for deriving the fault tolerant filter and fault detection equation in Section 4.

\subsection{Quantum Probability (Finite Dimensional Case)}
Let $(\Omega,\mathcal{F},\mathcal{P})$ be a complete classical probability space on which we have a right continuous and complete filtration $\{\mathcal{F}_t\}_{t\geq 0}$ of sub-$\sigma$ fields of $\mathcal{F}$. In the sequel, $\mathbb{E}_{\mathcal{P}}\{\cdot\}$ denotes the mathematical expectation operator with respect to the given probability measure $\mathcal{P}$.

We begin by introducing the quantum probability theory. Let $\mathcal{H}$ be a complex Hilbert space and $\mathscr{B}(\mathcal{H})$ be the set of all bounded operators on $\mathcal{H}$. We first discuss the case that $\dim(\mathcal{H})=n<\infty$. It is known that the foundations of quantum mechanics can be also formulated in a similar language to the classical Kolmogorov's probability theory (\cite{Gardiner2000}). The basic ideas are as follows. Based on the spectral theorem (\cite{Akhiezer1981}), any self-adjoint operator $A$ on $\mathcal{H}$ admits a spectral decomposition $A=\sum_{j=1}^n a_jP_{A_j}$, where $\{a_j\}\subset \mathbb{R}$ are the eigenvalues of $A$ and $\{P_{A_j}\}$ are the corresponding orthogonal projection operators which form a resolution of the identity, i.e., $P_{A_j}P_{A_k}=\delta_{jk}P_{A_k}$ and $\sum_{j=1}^n P_{A_j}=I$. For any continuous function $f: \mathbb{R}\to \mathbb{C}$, one has $f(A)=\sum_{j=1}^n f(a_j)P_{A_j}$. Thus the set $\mathscr{A}=\{X: X=f(A), f: \mathbb{R}\to \mathbb{C}\}$ forms a commutative $*-$algebra generated by $A$. That is, arbitrary linear combinations, products and adjoints of operators in $\mathscr{A}$ are still in $\mathscr{A}$, $I\in \mathscr{A}$ and all elements of $\mathscr{A}$ commute. A mapping $\mathbb{P}: \mathscr{A} \to \mathbb{C}$ is called a normal state on $\mathscr{A}$ if it is positive and normalized, i.e., $\mathbb{P}(X)\geq 0$ if $X\geq 0$ and $\mathbb{P}(I)=1$. From Theorem 7.1.12 in (\cite{Kadison1983}), there is always a density operator $\rho$ such that $\mathbb{P}(X)=\Tr(\rho X)$, where $\rho=\rho^{\dagger}, \Tr(\rho)=1$ and $\rho \geq 0$. Note that $P_{A_j}\in \mathscr{A}$ are exactly the events one can distinguish by measuring $A$ and their probabilities are given by $\mathbb{P}(A_j)$ if the system has a density operator $\rho$. We have the following lemma.

\textbf{Lemma 3.1} (\cite{Bouten2007}) Let $\mathscr{A}$ be a commutative $*-$algebra of operators on a finite-dimensional Hilbert space $\mathcal{H}$, and let $\mathbb{P}$ be a normal state on $\mathscr{A}$. There is a classical probability space $(\Omega',\mathcal{F}',\mathcal{P}')$ and a $*-$isomorphism\footnote[2]{A $*-$isomorphism $\iota$ is a linear bijection with $\iota(XY)=\iota(X)\iota(Y)$ and $\iota(X^{\dagger})=\iota(X)^{\dagger}$. Here $\iota$ depends only on a unitary operator $U$ by which all elements of the algebra $\mathscr{A}$ can be diagonalized. One can always find such an operator $U$ since all elements of $\mathscr{A}$ commute.} $\iota$ from $\mathscr{A}$ to the set of measurable functions on $\Omega'$, and moreover $\mathbb{P}(X)=\mathbb{E}_{\mathcal{P}'}(\iota(X)), \forall X\in \mathscr{A}$.

Thus a commutative $*-$algebra structure is equivalent to a classical probability space. The pair $(\{P_{A_j}\}, \mathbb{P})$ acts the same as $(\mathcal{F}',\mathcal{P}')$. An important conclusion from this isomorphic equivalence is that we are allowed to do fundamental mathematical manipulations on quantum observables and classical random variables in a similar way, i.e., if $X_1$ and $X_2$ are commuting self-adjoint operators that correspond to two classical random variables $x_1$ and $x_2$, respectively, then $X_1+X_2$ must correspond to $x_1+x_2$ and $X_1X_2$ must correspond to $x_1x_2$.What makes quantum probability model different from classical probability model is the existence of non-commutative observables. In classical probability, in every realization any event is either true or false, regardless of how many events we choose to observe and  the order of observations. However, in quantum probability, given a prior observation of an event $P$, any subsequent events that do not commute with $P$ become physically meaningless within the same realization. Consequently, joint statistics are only defined among commuting observables.

The quantum probability space is defined as follows.

\textbf{Definition 3.1} (\cite{Bouten2007}) A pair $(\mathscr{N}, \mathbb{P})$ is called a quantum probability space, where $\mathscr{N}$ is a $*-$algebra on $\mathcal{H}$.

\subsection{Joint Quantum-Classical Statistics}
In many physical situations quantum and classical randomnesses may coexist in system dynamics, which makes it desirable to define the joint quantum and classical statistics. Motivated by the systems described in Section 2, in the sequel we call observables in the following form ``random observables'':
\begin{equation}
A_R=\nu(R)U_R^{\dagger}AU_R. \label{gquantumAutoRevision1}
\end{equation}
Here $A$ is a self-adjoint operator on $\mathcal{H}$ representing any quantum observable; $R$ is a \emph{given} classical random vector defined on a classical probability space $(\Omega,\mathcal{F},\mathcal{P})$ and represents the classical random information in the quantum system dynamics. We suppose $R$ takes values in a finite set $\{R_1,...,R_{n_r}\}$; $U_R$ is a \emph{given} unitary operator-valued function of $R$ representing the random unitary evolution, i.e., $U_{R(\omega)}^{\dagger}U_{R(\omega)}\equiv I, \forall \omega \in \Omega$; $\nu(R)$ is a scalar function of $R$ representing a classical random variable of interest. Let $\mathscr{N}\subset \mathscr{B}(\mathcal{H})$ be a $*-$algebra as defined in Section 3.1. It follows from Section 7.2 in \cite{Bouten2009} that $A_R$ can be naturally considered to be an operator-valued random variable on a linear space $\ell^{\infty}(\Omega,\mathcal{F},\mathcal{P})\otimes \mathscr{N}$:
\begin{align}
A_R=\sum_{k=1}^{n_r}\nu(R_k)\textbf{1}_{R=R_k} \otimes U_{R_k}^{\dagger}AU_{R_k}\label{gquantumAutoRevision2}
\end{align}
where $\textbf{1}_{R=R_k}$ is the indicator function of the classical event ``$R=R_k$''. It is then clear that in each single measurement of the random observable $A_R$, we have to go through two realizations: (i) the choice of a sample point $\omega\in \Omega$, and (ii) the quantum measurement performed on a quantum observable $A_{R(\omega)}$. As a result, given a system state $\rho$, the average observed value of $A_R$ is denoted by $\tilde{\mathbb{P}}(A_R)$, where $\tilde{\mathbb{P}}$ is defined to be the linear mapping:
\begin{eqnarray}
\tilde{\mathbb{P}}(x\otimes X)=\mathbb{E}_{\mathcal{P}}\{x\Tr\{\rho X\}\}: \ell^{\infty}(\Omega,\mathcal{F},\mathcal{P})\otimes \mathscr{N}\to \mathbb{R},\label{gquantumAutoRevision3}
\end{eqnarray}
We refer to $\tilde{\mathbb{P}}$ as a quantum-classical expectation operator.

It is noted that random observables in the form of $A_R$ include any quantum observable of the form $U_R^{\dagger}AU_R$ and any classical random variable of the form $\nu(R)$ as special cases. Here, we treat any random variable $\nu(R)$ as a random observable $\nu(R)I$ under $\tilde{\mathbb{P}}$ because $\mathbb{E}_{\mathcal{P}}(e^{it\nu(R)})=\tilde{\mathbb{P}}(e^{it\nu(R)I})$ for any density operator $\rho$. In other words, $\nu(R)$ and $\nu(R)I$ are equivalent since they share the same characteristic function. It is clear that $\nu(R)I$ commutes with all quantum operators on $\mathcal{H}$ (this is exactly a property of classical random variables).

Define $\tilde{\mathscr{A}}$ to be a set of random observables $\tilde{\mathscr{A}}=\{X|X=\nu(R)f(U_R^{\dagger}AU_R), f: \mathbb{R}\to \mathbb{C}, \nu:\mathbb{R}^{n_r}\to \mathbb{C}\}$. It can be verified that for any functions $f_1, f_2 : \mathbb{R}\to \mathbb{C}$ and $\nu_1, \nu_2 :\mathbb{R}^{n_r}\to \mathbb{C}$, we have $[\nu_1(R)f_1(U_R^{\dagger}AU_R), \nu_2(R)f_2(U_R^{\dagger}AU_R)]=0$. That is, all elements in $\tilde{\mathscr{A}}$ commute.

Let $\mathscr{H}_0$ be a Hilbert space with $\dim\{\mathscr{H}_0\}=n_r$. Denote $\bar{\mathscr{H}}=\mathscr{H}_0\otimes \mathscr{H}$. The following result can be obtained with the proof presented in the Appendix.

\textbf{Theorem 3.1.} The set of random observables $\tilde{\mathscr{A}}$ equipped with the quantum-classical expectation operator $\tilde{\mathbb{P}}$ defined in (\ref{gquantumAutoRevision3}) is isomorphically equivalent to a quantum probability space $(\bar{\mathscr{R}}, \bar{\mathbb{P}})$, where $\bar{\mathscr{R}}$ is a commutative $*-$algebra on $\bar{\mathscr{H}}$, $\bar{\mathbb{P}}(X)=\Tr\{\bar \rho \otimes\rho X\}$ for any operator $X$ on $\bar{\mathscr{H}}$, and $\bar\rho$ is a density operator on $\mathscr{H}_0$.

\textbf{Remark 3.1.} From Theorem 3.1, any random observable can be equivalently represented by a quantum observable on a larger Hilbert space $\mathscr{H}_0\otimes \mathscr{H}$, which coincides with the way of describing a random observable in Definition 7.2 in \cite{Bouten2009}.

The following corollary can be directly concluded from Lemma 3.1 and Theorem 3.1.

\textbf{Corollary 3.1.} (General equivalence theorem, finite-dimensional case). There exist a probability space $(\Omega',\mathcal{F}',\mathcal{P}')$ and a $*-$isomorphism $\iota$ from $\tilde{\mathscr{A}}$ to the set of measurable functions on $\Omega'$, such that $\tilde{\mathbb{P}}(X)=\mathbb{E}_{\mathcal{P}'}(\iota(X)), \forall X\in \tilde{\mathscr{A}}$.

Thus the set $\tilde{\mathscr{A}}$ equipped with the quantum-classical expectation operator $\tilde{\mathbb{P}}$ is equivalent to a classical probability space. In other words, when the discussion is restricted to a set of commutative random observables, any probabilistic operation or joint statistics can be defined directly from the associated classical probability space. In particular, we consider the quantum-classical conditional expectation which will be used in subsequent analysis.

Let $Y_s\in \tilde{\mathscr{A}}'$ be a random observable, where $\tilde{\mathscr{A}}'=\{X|XY=YX, Y\in \tilde{\mathscr{A}}\}$ is the commutant of $\tilde{\mathscr{A}}$. Then $Y_s$ and $\tilde{\mathscr{A}}$ can generate a larger commutative set of random observables, which is isomorphic to a classical probability space through a linear mapping $\iota$ from Corollary 3.1. Following the same idea in classical probability theory, the map $\tilde{\mathbb{P}}(\cdot |\tilde{\mathscr{A}}): \tilde{\mathscr{A}}' \to \tilde{\mathscr{A}}$ is called (a version of) the conditional expectation from $\tilde{\mathscr{A}}'$ onto $\tilde{\mathscr{A}}$ if $\tilde{\mathbb{P}}(\tilde{\mathbb{P}}(X|\tilde{\mathscr{A}})Y)=\tilde{\mathbb{P}}(XY)$ for all $X\in \tilde{\mathscr{A}}', Y\in \tilde{\mathscr{A}}$, and a direct definition is given by $\tilde{\mathbb{P}}(Y_s|\tilde{\mathscr{A}})=\iota^{-1}(\mathbb{E}_{\mathcal{P}'}(\iota(Y_s)|\sigma\{\iota(\tilde{\mathscr{A}})\}))$.

From the spectral decomposition of $A$, one has
\begin{equation}
A_R=\sum_{j=1}^n\sum_{k=1}^{n_r}a_{jk}\tilde P_{jk},
\label{gquantumAutoRevision5}
\end{equation}
where $a_{jk}=a_j\nu(R_k)$ and $\tilde P_{jk}=\textbf{1}_{R=R_k} \otimes U_{R_k}^{\dagger}P_{A_j}U_{R_k}$. An explicit expression of the quantum-classical conditional expectation is given by
\begin{equation}
\tilde{\mathbb{P}}(X|\tilde{\mathscr{A}})=\sum \limits_{\tilde{\mathbb{P}}(\tilde P_{jk})\neq 0}\frac{\tilde{\mathbb{P}}(\tilde P_{jk}X)}{\tilde{\mathbb{P}}(\tilde P_{jk})}\tilde P_{jk}, \forall X\in \tilde{\mathscr{A}}'.\label{gquantum6}
\end{equation}
Here we investigate this expression further. Since $\tilde{\mathbb{P}}(X|\tilde{\mathscr{A}})\in \tilde{\mathscr{A}}$, by applying the $*-$isomorphism $\bar{\iota}=\iota_0\otimes I$ in Theorem 3.1 to both sides of (\ref{gquantum6}) we have
\begin{align}
\bar{\iota}\left(\tilde{\mathbb{P}}(X|\tilde{\mathscr{A}})\right)=\sum \limits_{\bar{\mathbb{P}}(\bar{\iota}(\tilde P_{jk})))\neq 0}\frac{\bar{\mathbb{P}}(\bar{\iota}(\tilde P_{jk})\bar{\iota}(X))}{\bar{\mathbb{P}}(\bar{\iota}(\tilde P_{jk}))}\bar{\iota}(\tilde P_{jk}), \label{gquantum6.1}
\end{align}
where $\bar{\iota}(\tilde P_{jk})=P_{R_k}\otimes U_{R_k}^{\dagger}P_{A_j}U_{R_k}$ from (\ref{gquantumAutoRevision4}). It follows from Theorem 3.1 that $\left\{\bar{\iota}(\tilde P_{jk})\right\}$ forms a set of basis projection operators for the commutative $*-$algebra $\bar{\mathscr{R}}$. Thus the expression (\ref{gquantum6}) is the same expression for quantum conditional expectation on $\bar{\mathscr{R}}$, as given in Equation (2.10) of (\cite{Bouten2007}). In fact, we have $\bar{\iota}\left(\tilde{\mathbb{P}}(X|\tilde{\mathscr{A}})\right)=\bar{\mathbb{P}}(\bar{\iota}(X)|\bar{\mathscr{R}})$.

Now consider the case $A\equiv I$, in which $\tilde{\mathscr{A}}$ is equivalent to the $\sigma-$field generated by the classical random variable $R$. Let $X=xI$ with $x$ being a random variable on $(\Omega,\mathcal{F},\mathcal{P})$. Then (\ref{gquantum6}) reduces to
\begin{eqnarray}
\tilde{\mathbb{P}}(X|\tilde{\mathscr{A}})=\sum \limits_{\mathbb{E}(\textbf{1}_{R=R_k})\neq 0}\frac{\mathbb{E}(x\textbf{1}_{R=R_k})}{\mathbb{E}(\textbf{1}_{R=R_k})}\textbf{1}_{R=R_k}=\mathbb{E}(x|\sigma\{R\}),\label{gquantum8}
\end{eqnarray}
which is the expression for classical conditional expectation (\cite{Bertsekas2002}).

Thus the defined conditional expectation is isomorphically equivalent to a particular quantum conditional expectation and contains classical conditional expectation as a special case. This coincides with the commonly accepted perspective that \emph{classical probability theory is a special case of quantum probability theory} \cite{Mirrahimi2007}. Note that Definition 3.1 also allows us to conveniently define the expectation of classical random variables conditioned on random observables, and vice versa.

The above analysis can be extended to the case when either $\Omega$ or $\mathscr{H}$ has infinite dimension. We will not give the details here. The key idea is that from Theorem 3.3 in (\cite{Bouten2007}) we can always construct on an additional Hilbert space a commutative von Neumann algebra which is isomorphic to the classical probability space $(\Omega,\mathcal{F},\mathcal{P})$. The overall linear space is thus isomorphic to the composition of two quantum probability spaces. Let $\mathscr{C}$ be a commutative von Neumann algebra on $\mathscr{H}$. Given a $\mathbb{R}^{n_r}$ valued classical random variable $R$ on $(\Omega,\mathcal{F},\mathcal{P})$ and a corresponding unitary operator $U_R$, define $\tilde{\mathscr{C}}=\{X|X=\nu(R)U_R^{\dagger}YU_R, Y\in \mathscr{C}, \nu:\mathbb{R}^{n_r}\to \mathbb{C}\}$ to be a set of commutative random observables equipped with the quantum-classical expectation operation $\tilde{\mathbb{P}}$. Here $\tilde{\mathbb{P}}$ is the same as that defined in (\ref{gquantumAutoRevision3}). From Theorem 3.3 in (\cite{Bouten2007}), one can prove that there exists a probability space $(\Omega',\mathcal{F}',\mathcal{P}')$ and a $*-$isomorphism $\iota$ from $\tilde{\mathscr{C}}$ onto the algebra of bounded measurable complex functions on $\Omega'$, such that $\tilde{\mathbb{P}}(X)=\mathbb{E}_{\mathcal{P}'}(\iota'(X)), X\in \tilde{\mathscr{C}}$. From classical probability theory, we have the following definition of quantum-classical conditional expectation.

\textbf{Definition 3.1.} (Quantum-classical conditional expectation) The map $\tilde{\mathbb{P}}(\cdot |\tilde{\mathscr{C}})$ is called (a version of) the quantum-classical conditional expectation from $\tilde{\mathscr{C}}'$ onto $\tilde{\mathscr{C}}$, if $\tilde{\mathbb{P}}(\tilde{\mathbb{P}}(X |\tilde{\mathscr{C}})Y)=\tilde{\mathbb{P}}(XY)$ for all $X\in \tilde{\mathscr{C}}'$ and $Y\in \tilde{\mathscr{C}}$.

It follows from Theorem 3.16 in (\cite{Bouten2007}) that the conditional expectation of Definition 3.1 exists and is unique with probability one (any two versions $P=\tilde{\mathbb{P}}(X |\tilde{\mathscr{C}})$ and $Q=\tilde{\mathbb{P}}(X |\tilde{\mathscr{C}})$ satisfy $\|P-Q\|_{\tilde{\mathbb{P}}}=0$, where $\|Y\|_{\tilde{\mathbb{P}}}=\tilde{\mathbb{P}}(Y^{\dagger}Y)$). Moreover, $\tilde{\mathbb{P}}(X|\tilde{\mathscr{C}})$ is the least mean square estimate of $X$ given $\tilde{\mathscr{C}}$ in the sense that $\|X-\tilde{\mathbb{P}}(X|\tilde{\mathscr{C}})\|\leq \|X-Y\|$ for all $Y\in \mathscr{C}$. One can verify that the elementary properties of classical conditional expectation, for example, linearity, positivity, the tower property and ``taking out what is known'' (\cite{Bertsekas2002}), still hold for the above defined conditional expectation in Definition 3.1.

In the subsequent application of fault tolerant quantum filtering we need to relate conditional expectations with respect to different states to each other. The following quantum-classical Bayes formula allows us to apply change of measure in both quantum and classical senses and is very useful in this problem.

\textbf{Theorem 3.2.} (Quantum-classical Bayes formula) Consider the classical probability space model $(\Omega,\mathcal{F},\mathcal{P})$, the set of random observables $\mathscr{C}$ and the quantum-classical expectation operator $\tilde{\mathbb{P}}$ defined as above. Suppose a new probability measure $\mathcal{Q}$ is defined by $d\mathcal{Q}=\Lambda d\mathcal{P}$, where the $\mathcal{F}-$measurable random variable $\Lambda$ is the classical Radon-Nikon derivative. Choose $V\in \tilde{\mathscr{C}}'$ such that $V^{\dagger}V>0$ and $\tilde{\mathbb{P}}(\Lambda V^{\dagger}V)=1$. Then we can define on $\tilde{\mathscr{C}}'$ a new quantum-classical expectation operator $\tilde{\mathbb{Q}}$ by $\tilde{\mathbb{Q}}(X)=\tilde{\mathbb{P}}(\Lambda V^{\dagger}XV)$ and
\begin{equation}
\tilde{\mathbb{Q}}(X|\tilde{\mathscr{C}})=\frac{\tilde{\mathbb{P}}(\Lambda V^{\dagger}XV/\tilde{\mathscr{C}})}{\tilde{\mathbb{P}}(\Lambda V^{\dagger}V/\tilde{\mathscr{C}})}, \hspace{1cm}\forall X\in \tilde{\mathscr{C}}'. \label{gquantum9}
\end{equation}
\emph{Proof.}  Let $Y$ be any element of $\tilde{\mathscr{C}}$. Then we have
\begin{eqnarray}
&&\tilde{\mathbb{P}}(\tilde{\mathbb{P}}(\Lambda V^{\dagger}XV|\tilde{\mathscr{C}})Y)=\tilde{\mathbb{P}}(\Lambda V^{\dagger}XVY)\nonumber\\
&=&\tilde{\mathbb{P}}(\Lambda V^{\dagger}XYV)\nonumber\\
&=&\tilde{\mathbb{Q}}(XY)=\tilde{\mathbb{Q}}(\tilde{\mathbb{Q}}(XY|\tilde{\mathscr{C}}))\nonumber\\
&=&\tilde{\mathbb{P}}(\Lambda V^{\dagger}\tilde{\mathbb{Q}}(X|\tilde{\mathscr{C}})YV)=\tilde{\mathbb{P}}(\Lambda V^{\dagger}V\tilde{\mathbb{Q}}(X|\tilde{\mathscr{C}})Y)\nonumber\\
&=&\tilde{\mathbb{P}}(\tilde{\mathbb{P}}(\Lambda V^{\dagger}V|\tilde{\mathscr{C}})\tilde{\mathbb{Q}}(X|\tilde{\mathscr{C}})Y). \label{Bayes}
\end{eqnarray}
Let $Y=(\tilde{\mathbb{P}}(\Lambda V^{\dagger}XV|\tilde{\mathscr{C}})-\tilde{\mathbb{P}}(\Lambda V^{\dagger}V|\tilde{\mathscr{C}})\tilde{\mathbb{Q}}(X|\tilde{\mathscr{C}}))^{\dagger}$, then from (\ref{Bayes}) we have $\|Y\|_{\tilde{\mathbb{P}}}=0$. In other words, $\tilde{\mathbb{P}}(\Lambda V^{\dagger}XV|\tilde{\mathscr{C}})=\tilde{\mathbb{P}}(\Lambda V^{\dagger}V|\tilde{\mathscr{C}})\tilde{\mathbb{Q}}(X|\tilde{\mathscr{C}})$ $\tilde{\mathbb{P}}$ almost surely. $\hspace{2cm}\qed$

\textbf{Remark 3.2.} Theorem 3.2 is equivalent to the quantum Bayes formula (\cite{Bouten2007}) and contains classical Bayes formula (\cite{Bertsekas2002}) as a special case.

\section{Fault Tolerant Quantum Filtering and Fault Detection}

\subsection{Fault tolerant quantum filter and fault detection equation}
In classical (non-quantum) engineering, apparatuses may suffer from malfunctions or degradation events (faults), especially after a long running time or when working in difficult environments. The occurrence of faults can often make the system evolve far from its desired or normal operating conditions and can lead to a drastic change in the system behaviour. Thus this is a phenomenon that needs to be seriously considered. Recall the quantum systems described in Section 2. In the laser-atom interaction realization, the laser field is often treated in a classical way and it generates an electromagnetic field at the position of the atom. Then the laser-atom interaction can be described by a dipole interaction Hamiltonian which depends on the intensity of the classical electromagnetic field (\cite{Ruschhaupt2012}). Therefore, if the macroscopic laser device suffers from a fault, e.g., it produces a faulty electromagnetic field, an unexpected additional Hamiltonian will be introduced into the quantum system. In this case, the system Hamiltonian in (\ref{gquantum2}) will be given by $H(F(t))$ where $F(t)$ is the fault process. 

In practice, the system may transit between a finite number of different faulty modes at random times. This makes it desirable to model the fault process on a probability space $(\Omega,\mathcal{F},\mathcal{P})$ by a continuous-time Markov chain $\{F(t)\}_{t\geq 0}$ adapted to $\{\mathcal{F}_t\}_{t\geq 0}$ (\cite{Davis1975}, \cite{Hibey1999},
\cite{Elliott1995}). The state space of $F(t)$ is often chosen to be the finite set $\mathbb{S}=\{e_1, e_2,...,e_N\}$ (for some positive integer $N$) of canonical unit vectors in $\mathbb{R}^N$. Let $p_t=(p_t^1, p_t^2,...,p_t^N)^T$ be the probability distribution of $F(t)$, i.e., $p_t^k=\mathcal{P}(F(t)=e_k), k=1,2,...,N$ and suppose the Markov process $F(t)$ has a so-called Q matrix or transition rate matrix $\Pi=(a_{jk})\in \mathbb{R}^{N\times N}$. Then $p_t$ satisfies the forward Kolmogorov equation $ \frac{dp_t}{dt}=\Pi p_t$. Because $\Pi$ is a Q matrix, we have $a_{jj}=-\sum_{j\neq k}a_{jk}$, and $a_{jk}\geq 0, j\neq k$. Then $F(t)$ is a corlol process (\cite{Elliott1995}) that satisfies the following stochastic differential equation:
\begin{equation}
dF(t)=\Pi F(t)dt+dM(t), \label{gquantum1}
\end{equation}
where $M(t)=F(t)-F(0)-\int_0^t\Pi F(\tau^-)d\tau$ is an $\{\mathcal{F}_t\}$ martingale (\cite{Elliott1995}) and satisfies
\begin{equation*}
\sup \limits_{0\leq t \leq T} \mathbb{E}(|M(t)|^2)< \infty.
\end{equation*}
One goal of this paper is to derive the equations of the fault tolerant quantum filter and fault detection for this class of open quantum systems. To be specific, we use a reference probability approach to find the least-mean-square estimates of a system observable $X\in \mathscr{B}(\mathcal{H})$ at time $t$ and the fault process $F(t)$ for the quantum system under consideration, given the observation process $Y(s), 0\leq s \leq t$. This can be accomplished if we can obtain the following estimates:
\begin{equation}
\sigma_t^j(X)=\tilde{\mathbb{P}}(\left<F(t),e_j\right>U^{\dagger}(t)XU(t)|\mathscr{Y}_t),\label{gquantum10}
\end{equation}
where $\mathscr{Y}_t$ is the commutative von Neumann algebra generated by $Y(s)$ up to time $t$, and $\left<\cdot, \cdot\right>$ is the inner product in $\mathbb{R}^N$. From the previous analysis, one has $\left<F(t),e_j\right>U^{\dagger}(t)XU(t)\in \mathscr{Y}_t'$, which guarantees that the conditional expectation (\ref{gquantum10}) is well defined.

It follows from (\ref{gquantum4}) that for $\forall s \leq t$,
\begin{eqnarray}
U^{\dagger}(t)Q(s)U(t)&=&U^{\dagger}(s)U^{\dagger}(t,s)Q(s)U(t,s)U(s)\nonumber\\
&=&U^{\dagger}(s)Q(s)U(s)=Y(s), \label{gquantum11}
\end{eqnarray}
which implies that $\mathscr{Y}_t$ can be rewritten as $\mathscr{Y}_t=U^{\dagger}(t)\mathscr{Q}_tU(t)$ where $\mathscr{Q}_t$ is the commutative von Neumann algebra generated by $Q(s)$ up to time $t$. From quantum probability theory, we know that $Q(t)$ under the vacuum state is equivalent to a classical Wiener process (\cite{Gardiner2000}). This fact makes it simpler to design a quantum filter in terms of $Q(t)$ because it is convenient to manipulate $Q(t)$ using the quantum $It\hat o$ formula (\cite{Hudson1984}). Next, we will use a quantum analog of the classical change-of-measure technique to obtain an explicit expression for $\sigma_t^j(X)$.

Define an operator $V(t)$ that satisfies the quantum stochastic differential equation
\begin{equation}
dV(t)=\left\{\left(-iH(F(t))-\frac{1}{2}L^{\dagger}L\right)dt+LdQ(t)\right\}V(t),\label{gquantum12}
\end{equation}
with $V(0)=I$. Then $V(t)\in \mathscr{Q}_t'$ and we have the following lemma.

\textbf{Lemma 4.1.}  For any system observable $X\in \mathscr{B}(\mathcal{H})$, the conditional expectation in (\ref{gquantum10}) can be rewritten as
\begin{equation}
\sigma_t^j(X)=U^{\dagger}(t)\frac{\tilde{\mathbb{P}}(\left<F(t),e_j\right>V^{\dagger}(t)XV(t)|\mathscr{Q}_t)}{\tilde{\mathbb{P}}(V^{\dagger}(t)V(t))|\mathscr{Q}_t)}U(t). \label{gquantum13}
\end{equation}
\emph{Proof.} See the Appendix. $\hspace{4.3cm}\qed$

Write 
\begin{equation}
\pi_t^j(X)=U^{\dagger}(t)\tilde{\mathbb{P}}(\left<F(t),e_j\right>V^{\dagger}(t)XV(t)|\mathscr{Q}_t)U(t),  \label{gquantum14}
\end{equation}
which is the unnormalized conditional expectation. Since $\sum_{j=1}^N\left<F(t),e_j\right>=1$, we have
\begin{equation}
\sigma_t^j(X)=\frac{\pi_t^j(X)}{\sum_{k=1}^{N}\pi_t^k(I)}. \label{gquantum15}
\end{equation}
An explicit expression for $\pi_t^j(X)$ can now be obtained.

\textbf{Theorem 4.1.} (Unnormalized fault tolerant quantum filtering equation) The unnormalized conditional expectation $\pi_t^j(X)$ satisfies the following quantum stochastic differential equation:
\begin{eqnarray}
d\pi_t^j(X)&=&\left(\sum_{k=1}^Na_{jk}\pi_t^k(X)+\pi_t^j(\mathscr{L}_{L, H(e_j)}(X))\right)dt\nonumber\\
&&+\pi_t^j(XL+L^{\dagger}X)dY(t), \label{gquantum16}
\end{eqnarray}
where the so-called Lindblad generator is given by
\begin{equation*}
\mathscr{L}_{L, H}(X)=i[H,X]+L^{\dagger}XL-\frac{1}{2}(L^{\dagger}LX+XL^{\dagger}L). 
\end{equation*}
\emph{Proof.} Using the $It\hat o$ product rule, and from (\ref{gquantum1}) and (\ref{gquantum12}), we obtain
\begin{eqnarray}
&&\left<F(t),e_j\right>V^{\dagger}(t)XV(t)\nonumber\\
&=&\left<F(0),e_j\right>X+\int_0^t \left<\Pi F(s),e_j\right>V^{\dagger}(s)XV(s)ds\nonumber\\
&&+\left<\int_0^tV^{\dagger}(s)XV(s)dM(s),e_j\right>\nonumber\\
&&+\int_0^t \left<F(s),e_j\right>d(V^{\dagger}(s)XV(s)). \label{gquantum18}
\end{eqnarray}
Taking conditional expectation with respect to $\mathscr{Q}_t$ on both sides of (\ref{gquantum18}) while using the mutual independence of $\{Q(t), M(t), F(0)\}$, we obtain
\begin{eqnarray}
&&\tilde{\mathbb{P}}(\left<F(t),e_j\right>V^{\dagger}(t)XV(t)|\mathscr{Q}_t)\nonumber\\
&=&\tilde{\mathbb{P}}(\left<F(0),e_j\right>X)\nonumber\\
&&+\tilde{\mathbb{P}}\left(\int_0^t \left<\Pi F(s),e_j\right>V^{\dagger}(s)XV(s)ds|\mathscr{Q}_t\right)\nonumber\\
&&+\tilde{\mathbb{P}}\left(\int_0^t \left<F(s),e_j\right>V^{\dagger}(s)\mathscr{L}_{L, H(F(s))}(X)V(s)ds|\mathscr{Q}_t\right)\nonumber\\
&&+\tilde{\mathbb{P}}\left(\int_0^t \left<F(s),e_j\right>V^{\dagger}(s)(XL+L^{\dagger}X)V(s)dQ(s)|\mathscr{Q}_t\right)\nonumber\\
&=&\tilde{\mathbb{P}}(\left<F(0),e_j\right>X)\nonumber\\
&&+\int_0^t \tilde{\mathbb{P}}(\left<\Pi F(s),e_j\right>V^{\dagger}(s)XV(s)|\mathscr{Q}_s)ds\nonumber\\
&&+\int_0^t \tilde{\mathbb{P}}\left(\left<F(s),e_j\right>V^{\dagger}(s)\mathscr{L}_{L, H(e_j)}(X)V(s)|\mathscr{Q}_s\right)ds\nonumber\\
&&+\int_0^t \tilde{\mathbb{P}}\left( \left<F(s),e_j\right>V^{\dagger}(s)(XL+L^{\dagger}X)V(s)|\mathscr{Q}_s\right)dQ(s).\label{gquantum19}
\end{eqnarray}
In addition,
\begin{eqnarray}
\left<\Pi F(s),e_j\right>&=&\left<F(s),\Pi^{T}e_j\right>=\left<F(s),\sum_{k=1}^N a_{jk}e_k\right>\nonumber\\
&=&\sum_{k=1}^N a_{jk}\left<F(s),e_k\right>. \label{gquantum20}
\end{eqnarray}
Let $h_t^j(X)=\tilde{\mathbb{P}}(\left<F(t),e_j\right>V^{\dagger}(t)XV(t)|\mathscr{Q}_t)$. Then we have $\pi_t^j(X)=U^{\dagger}(t)h_t^j(X)U(t)$. From (\ref{gquantum19}) and (\ref{gquantum20}), $h_t^j(X)$ satisfies the following stochastic differential equation:
\begin{eqnarray}
dh_t^j(X)&=&\left(\sum_{k=1}^N a_{jk} h_t^k(X)+h_t^j\left(\mathscr{L}_{L, H(e_j)}(X)\right)\right)dt\nonumber\\
&&+h_t^j(XL+L^{\dagger}X)dQ(t).\label{gquantum21}
\end{eqnarray}
From Definition 3.1, we know $h_t^j(X) \in \mathscr{Q}_t$. Using the $It \hat o$ formula, we have
\begin{eqnarray}
d\pi_t^j(X)&=&(U(t)+dU(t))^{\dagger}dh_t^j(X)(U(t)+dU(t)) \nonumber\\
&=&\left(\sum_{k=1}^N a_{jk} \pi_t^k(X)+\pi_t^j(\mathscr{L}_{L, H(e_j)}(X))\right)dt\nonumber\\
&&+\pi_t^j(XL+L^{\dagger}X)dQ(t)\nonumber\\
&&+\pi_t^j(XL+L^{\dagger}X)U^{\dagger}(t)(L+L^{\dagger})U(t) \nonumber\\
&=&\left(\sum_{k=1}^N a_{jk} \pi_t^k(X)+\pi_t^j(\mathscr{L}_{L, H(e_j)}(X))\right)dt\nonumber\\
&&+\pi_t^j(XL+L^{\dagger}X)dY(t), \label{gquantum22}
\end{eqnarray}
which is exactly (\ref{gquantum16}).$\hspace{4.5cm}\qed$

\textbf{Theorem 4.2.} (Normalized fault tolerant quantum filtering equation) The normalized conditional expectation $\sigma_t^j(X)$ satisfies the following quantum stochastic differential equation:
\begin{eqnarray}
d\sigma_t^j(X)=(\sum_{k=1}^Na_{jk}\sigma_t^k(X)+\sigma_t^j(\mathscr{L}_{L, H(e_j)}(X)))dt+\nonumber\\
\left(\sigma_t^j(XL+L^{\dagger}X)-\sigma_t^j(X)\sum_{k=1}^N\sigma_t^k(L+L^{\dagger})\right)dW(t), \hspace{0.5cm} \label{gquantum23}
\end{eqnarray}
where $W(t)=Y(t)-\int_0^t \sum_{k=1}^N\sigma_s^k(L+L^{\dagger})ds$ is called innovation process and is a Wiener process under $\tilde{\mathbb{P}}$.

\emph{Proof.} From Theorem 4.1, we have
\begin{eqnarray}
d\pi_t^j(I)=\sum_{k=1}^Na_{jk}\pi_t^k(I)dt+\pi_t^j(L+L^{\dagger})dY(t), \label{gquantum24}
\end{eqnarray}
since $\mathscr{L}_{L, H(e_j)}(I)=0$.

In addition, it follows from the properties of the Q matrix that
\begin{eqnarray}
d\sum_{k=1}^N \pi_t^k(I)&=&\sum_{j=1}^N\sum_{k=1}^Na_{jk}\pi_t^k(I)dt+\sum_{k=1}^N\pi_t^k(L+L^{\dagger})dY(t)\nonumber\\
&=&\sum_{k=1}^N\pi_t^k(L+L^{\dagger})dY(t). \label{gquantum25}
\end{eqnarray}
Equation (\ref{gquantum15}) can be rewritten as
\begin{equation}
\sum_{k=1}^{N}\pi_t^k(I)\sigma_t^j(X)=\pi_t^j(X). \label{gquantum26}
\end{equation}
Differentiating both sides of (\ref{gquantum26}) based on the quantum $It\hat o$ rule yields
\begin{equation}
d\sum_{k=1}^{N}\pi_t^k(I)(\sigma_t^j(X)+d\sigma_t^j(X))+\sum_{k=1}^{N}\pi_t^k(I)d\sigma_t^j(X)
=d\pi_t^j(X).\label{gquantum27}
\end{equation}
It is noted that $[\sigma_t^j(X), dY(t)]=0$ because $\sigma_t^j(X)\in \mathscr{Y}_t$. From (\ref{gquantum24})-(\ref{gquantum27}), one has
\begin{eqnarray}
&&\left(\sum_{k=1}^{N}\pi_t^k(I)+\sum_{k=1}^N\pi_t^k(L+L^{\dagger})dY(t)\right)d\sigma_t^j(X)\nonumber\\
&=&d\pi_t^j(X)-\sum_{k=1}^N\pi_t^k(L+L^{\dagger})\sigma_t^j(X)dY(t).\label{gquantum28}
\end{eqnarray}
From (\ref{gquantum16}) and (\ref{gquantum26}), one has
\begin{eqnarray}
&&\left(\sum_{k=1}^{N}\pi_t^k(I)\right)^{-1}d\pi_t^j(X)\nonumber\\
&=&\left(\sum_{k=1}^Na_{jk}\sigma_t^k(X)+\sigma_t^j(\mathscr{L}_{L, H(e_j)}(X))\right)dt\nonumber\\
&&+\sigma_t^j(XL+L^{\dagger}X)dY(t). \label{gquantum29}
\end{eqnarray}
Then dividing both sides of (\ref{gquantum28}) by $\sum_{k=1}^{N}\pi_t^k(I)$ yields
\begin{eqnarray}
\left(I+\sum_{k=1}^N\sigma_t^k(L+L^{\dagger})dY(t)\right)d\sigma_t^j(X)\hspace{2.4cm}\nonumber\\
=\left(\sum_{k=1}^Na_{jk}\sigma_t^k(X)+\sigma_t^j(\mathscr{L}_{L, H(e_j)}(X))\right)dt\hspace{2cm}\nonumber\\
+\left(\sigma_t^j(XL+L^{\dagger}X)-\sum_{k=1}^N\sigma_t^k(L+L^{\dagger})\sigma_t^j(X)\right)dY(t).\hspace{4mm}\label{gquantum30}
\end{eqnarray}
By multiplying both sides of (\ref{gquantum30}) with $I-\sum_{k=1}^N\sigma_t^k(L+L^{\dagger})dY(t)$, (\ref{gquantum23}) can be obtained using the fact $dY(t)dY(t)=dt$.

Next, note $\sum_{k=1}^N\sigma_t^k(L+L^{\dagger})=\tilde{\mathbb{P}}(U^{\dagger}(t)(L+L^{\dagger})U(t)|\mathscr{Y}_t)\in \mathscr{Y}_t$. Thus one can prove that $W(t)$ is a commutative process which is equivalent to a classical stochastic process under $\tilde{\mathbb{P}}$ according to Corollary 3.1.

In addition, let $K\in \mathscr{Y}_s, s\leq t$, then
\begin{eqnarray*}
&&\tilde{\mathbb{P}}(\tilde{\mathbb{P}}(W(t)|\mathscr{Y}_s)K)=\tilde{\mathbb{P}}(W(t)K)\nonumber\\
&=&\tilde{\mathbb{P}}\left(Y(t)K-\int_0^t \tilde{\mathbb{P}}(U^{\dagger}(\tau)(L+L^{\dagger})U(\tau)|\mathscr{Y}_{\tau}))Kd\tau\right)\nonumber\\
\end{eqnarray*}
\begin{eqnarray}
&=&\tilde{\mathbb{P}}\left(Y(t)K-\int_0^s \tilde{\mathbb{P}}(U^{\dagger}(\tau)(L+L^{\dagger})U(\tau)|\mathscr{Y}_{\tau}))Kd\tau\right.\nonumber\\
&&\left.-\int_s^t U^{\dagger}(\tau)(L+L^{\dagger})U(\tau)d\tau K\right)\nonumber\\
&=&\tilde{\mathbb{P}}(W(s)K)+\tilde{\mathbb{P}}((Q(t)-Q(s))K)=\tilde{\mathbb{P}}(W(s)K).\label{gquantum31}
\end{eqnarray}
Therefore, $\tilde{\mathbb{P}}(W(t)|\mathscr{Y}_s)=W(s), s\leq t$, which means $W(t)$ is a $\mathscr{Y}_t-$martingale.  Finally, $dW(t)dW(t)=dY(t)dY(t)=dt$. Then $W(t)$ is a Wiener process using Levy's Theorem (\cite{Karatsas1991}). $\hspace{3.1cm}\qed$

\textbf{Remark 4.1.}  Since our discussion is under the Heisenberg picture, $\tilde{\mathbb{P}}$ is fixed. Based on Corollary 3.1, (\ref{gquantum23}) is a classical recursive stochastic differential equation driven by the classical Wiener process $W(t)$, and $Y(t)$ can be replaced by its classical observation process counterpart. As a result, (\ref{gquantum23}) can be directly implemented on a classical signal processor.

\textbf{Remark 4.2.} The coupled system of stochastic differential equations (\ref{gquantum23}) is the normalized conditional expectation of $\left<F(t),e_j\right>U^{\dagger}(t)XU(t)$, given $\mathscr{Y}_t$. When $\pi_{jk}=0, \forall j\neq k$, this system is decoupled and reduces to the well known quantum filtering equation of $U^{\dagger}(t)XU(t)$ given $\mathscr{Y}_t$ (\cite{Belavkin1992}, \cite{Bouten2007}).

Normally, the open quantum system is defined on a finite dimensional Hilbert space $\mathcal{H}_s$. Noting that $\sigma_t^j$ is a linear, identity preserving and positive mapping on $\mathscr{Y}_t'$. From another point of view, it works as the expectation of $\left<F(t),e_j\right>X$ with respect to some finite dimensional state on $\mathcal{H}_s$. Thus there exists a density operator $\rho_t'$ such that $\sigma_t^j(X)=\mathbb{E}\{\Tr\{\rho_t' (\left<F(t),e_j\right>X)\}\}=\Tr\{\rho_t^jX\}$ with $\rho_t^j=\mathbb{E}\left(\left<F(t),e_j\right>\rho_t'\right)$. The following is a corollary of Theorem 4.2.

\textbf{Corollary 4.1.}  Let $\rho_t^j$ be the random operator that satisfies $\sigma_t^j(X)=\Tr(\rho_t^jX)$ for all system observables $X\in \mathscr{B}(\mathcal{H})$. Then $\rho_t^j$ satisfies the following stochastic differential equation
\begin{eqnarray}
&&d\rho_t^j=\left(\sum_{k=1}^Na_{jk}\rho_t^k+\mathscr{L}_{L, H(e_j)}^{\dagger}(\rho_t^j)\right)dt\nonumber\\
&&\hspace{0.6cm}+\left(L\rho_t^j+\rho_t^jL^{\dagger}-\rho_t^j\sum_{k=1}^N\Tr(\rho_t^k(L+L^{\dagger}))\right)dW(t), \hspace{0.5cm} \label{gquantum32}
\end{eqnarray}
with $\rho_0^j=\mathbb{E}(\left<F(0),e_j\right>)\pi_0$. Here $\mathscr{L}_{L, H(e_j)}^{\dagger}$ is the adjoint Lindblad generator:
\begin{equation*}
\mathscr{L}_{L, H}^{\dagger}(X)=-i[H,X]+LXL^{\dagger}-\frac{1}{2}(L^{\dagger}LX+XL^{\dagger}L). 
\end{equation*}
Note $\rho_t^j$ is not a density matrix because it is not defined in terms of the conditional expectation of real system observables. In fact, we have
\begin{equation}
\tilde{\mathbb{P}}(U^{\dagger}(t)XU(t)|\mathscr{Y}_t)=\sum_{k=1}^N\sigma_t^k(X).\label{gquantum34}
\end{equation}
Let $\rho_t$ be the random density matrix that satisfies $\tilde{\mathbb{P}}(U^{\dagger}(t)XU(t)|\mathscr{Y}_t)=\Tr(\rho_tX)$. We have
\begin{equation}
\rho_t=\sum_{k=1}^N\rho_t^k, \mbox{ with } \Tr(\rho_t)=1 \mbox{ and } \rho_0=\pi_0.\label{gquantum35}
\end{equation}
From Corollary 4.1, $\rho_t$ satisfies
\begin{eqnarray}
d\rho_t&=&\left(-\sum_{k=1}^Ni[H(e_k),\rho_t^k]+L\rho_tL^{\dagger}-\frac{1}{2}L^{\dagger}L\rho_t-\frac{1}{2}\rho_tL^{\dagger}L\right)dt\nonumber\\
&&+(L\rho_t+\rho_tL^{\dagger}-\rho_t\Tr((L+L^{\dagger})\rho_t)dW(t).\label{gquantum36}
\end{eqnarray}
Equation (\ref{gquantum36}) is the \emph{fault tolerant quantum stochastic master equation}.

In addition, the conditional probability densities of the fault process are given by
\begin{equation}
\hat p_t^j=\mathcal{P}(F(t)=e_j|\mathscr{Y}_t)=\tilde{\mathbb{P}}(\left<F(t),e_j\right>|\mathscr{Y}_t)=\sigma_t^j(I), \label{revise1}
\end{equation}
which satisfy the following coupled equations using Theorem 4.2:
\begin{eqnarray}
d\hat p_t^j&=&\sum_{k=1}^Na_{jk}\hat p_t^kdt\nonumber\\
&&+\left(\sigma_t^j(L+L^{\dagger})-\hat p_t^j\sum_{k=1}^N\sigma_t^k(L+L^{\dagger})\right)dW(t). \hspace{0.5cm} \label{revise2}
\end{eqnarray}
Let $\hat p_t=[\hat p_t^1,...,\hat p_t^N]'$. Then (\ref{revise2}) can be rewritten in a vector form as
\begin{equation}
d\hat p_t=\Pi\hat p_tdt+G(t)dW(t),\label{revise3}
\end{equation}
where $G(t)=\sum_{k=1}^Ne_k\sigma_t^k(L+L^{\dagger})-\hat p_t\sum_{k=1}^N\sigma_t^k(L+L^{\dagger})$.
Equation (\ref{revise3}) is the corresponding \emph{fault detection equation}.

The system of coupled equations (\ref{revise2}) or the vector form (\ref{revise3}) represents the conditional probability distribution that the system is under any faulty mode. It can be used to determine whether a particular type of fault has happened within the system at time $t$. A possible criteria for fault detection is 
\begin{equation}
\mbox{The $j$th fault happens, if } \hat p_t^j\geq p_0,
\end{equation}
where $1\geq p_0>0$ is a threshold value chosen by the users.


\newcounter{TempEqCnt}
\setcounter{TempEqCnt}{\value{equation}} \setcounter{equation}{47}
\begin{figure*}[ht!]
\begin{equation}
\left\{
\begin{array}{l}
d\alpha(t)=-\lambda\alpha(t)dt+\frac{1}{\sqrt{T_1}}(x_1(t)-\alpha(t)(x_1(t)+x_2(t)))dW(t)\\
dx_1(t)=-((\lambda+\frac{1}{2T_1})x_1(t)+\omega_z y_1(t))dt+\frac{1}{\sqrt{T_1}}(\alpha(t)+z_1(t)-x_1(t)(x_1(t)+x_2(t)))dW(t)\\
dy_1(t)=(\omega_z x_1(t)-(\lambda+\frac{1}{2T_1})y_1(t))dt-\frac{1}{\sqrt{T_1}}(x_1(t)+x_2(t))y_1(t)dW(t)\\
dz_1(t)=-(\frac{1}{T_1}\alpha(t)+(\lambda+\frac{1}{T_1})z_1(t))dt-\frac{1}{\sqrt{T_1}}(x_1(t)+(x_1(t)+x_2(t))z_1(t))dW(t)\\
dx_2(t)=(\lambda x_1(t)-\frac{1}{2T_1}x_2(t)-\omega_zy_2(t)+\omega_yz_2(t))dt+\frac{1}{\sqrt{T_1}}(1-\alpha(t)-x_2(t)(x_1(t)+x_2(t))+z_2(t))dW(t)\\
dy_2(t)=(\lambda y_1(t)+\omega_z x_2(t)-\frac{1}{2T_1}y_2(t))dt-\frac{1}{\sqrt{T_1}}(x_1(t)+x_2(t))y_2(t)dW(t)\\
dz_2(t)=(\lambda z_1(t)-\omega_yx_2(t)-\frac{1}{T_1}(1-\alpha(t)+z_2(t)))dt-\frac{1}{\sqrt{T_1}}(x_2(t)+(x_1(t)+x_2(t))z_2(t))dW(t)
\end{array}
\right. \label{filtering}
\end{equation}
\end{figure*}
\setcounter{equation}{\value{TempEqCnt}}

\subsection{Application to Two-level Quantum Systems}
Two-level quantum systems (qubits) play a significant role in quantum information processing. For a two-level system, the filter equations reduce to a finite set of stochastic differential equations. In this case, $\mathcal{H}_s=\mathbb{C}^2$. Denote the Pauli matrices by
$\sigma_x=
\left(
\begin{array}{cc}
  0  & 1   \\
 1   & 0  
\end{array}
\right), \sigma_y=
\left(
\begin{array}{cc}
  0  & -i   \\
 i   & 0  
\end{array}
\right) \mbox{ and }\sigma_z=
\left(
\begin{array}{cc}
  1  & 0   \\
 0   & -1  
\end{array}
\right).$
We select the coupling strength operator $L=\sqrt{1/T_1}\sigma_{-}$ and the free Hamiltonian $H_0=\frac{\omega_z}{2}\sigma_z$, where $T_1$ is the life time of the excited state, $\sigma_{-}=\frac{1}{2}(\sigma_x-i\sigma_y)$ and $\omega_z$ is the two-level pulsation.

Assume that a fault occurs at time $T$, at which time a new Hamiltonian $H_f=\frac{\omega_y}{2}\sigma_y$ is introduced into the system, where $\omega_y$ is an additional pulsation. Following (\cite{Davis1975}), we assume that $f(t)$ is a Poisson process with rate $\lambda$, stopped at its first jump time $T$. That is,
\begin{equation}
f(t)=
\left\{
\begin{array}{c}
0, \mbox{ if } t<T\\
1, \mbox{ if } t\geq T
\end{array}
\right.
\end{equation}
and $T$ is an exponential random variable with probability distribution
\begin{equation}
P(T\leq t)=1-e^{-\lambda t}.
\end{equation}
From (\cite{Elliott1995}), the process $M(t)=f(t)-\lambda \min(t,T)$ is a martingale and the process $f(t)$ satisfies
\begin{equation}
df(t)=\lambda(1-f(t))dt+dM(t).
\end{equation}
Also, we consider $f(0)=0$ only (because $f(t)$ stops at its first jump). Let $F(t)=[1-f(t),f(t)]'$. Then $F(t)$ takes values in $\{e_1,e_2\}$ and satisfies
\begin{equation}
dF(t)=\left[
\begin{array}{cc}
-\lambda & 0\\
\lambda  & 0
\end{array}
\right]F(t)+\left[
\begin{array}{c}
-1\\
1
\end{array}
\right]dM(t).
\end{equation}
Hence, the coupled quantum filtering equations are given by
\begin{equation}
\left\{
\begin{array}{l}
d\rho_t^1=\left(-\lambda\rho_t^1+\mathscr{L}_{L, H_0}^{\dagger}(\rho_t^1)\right)dt\nonumber\\
+\left(L\rho_t^1+\rho_t^1L^{\dagger}-\rho_t^1\sum_{k=1}^2\Tr(\rho_t^k(L+L^{\dagger}))\right)dW(t), \\
d\rho_t^2=\left(\lambda\rho_t^1+\mathscr{L}_{L, H_0+H_f}^{\dagger}(\rho_t^2)\right)dt\nonumber\\
+\left(L\rho_t^2+\rho_t^2L^{\dagger}-\rho_t^2\sum_{k=1}^2\Tr(\rho_t^k(L+L^{\dagger}))\right)dW(t). 
\end{array}
\right.
\end{equation}
Write 
\begin{eqnarray*}
\left\{
\begin{array}{l}
\rho_t^1=\frac{1}{2}(\alpha(t)I+x_1(t)\sigma_x+y_1(t)\sigma_y+z_1(t)\sigma_z),\\
\rho_t^2=\frac{1}{2}((1-\alpha(t))I+x_2(t)\sigma_x+y_2(t)\sigma_y+z_2(t)\sigma_z).
\end{array}
\right.
\end{eqnarray*}
Then we obtain seven coupled equations for the seven coefficients related to the fault tolerant quantum stochastic master equation in (\ref{filtering}) at the top on this page.

The fault detection equation is given by
\setcounter{equation}{48}
\begin{equation}
d\hat p_t=\Pi \hat p_tdt+\frac{1}{\sqrt{T_1}}G(t)dW(t).
\end{equation}
where $G(t)=\left(\begin{array}{c}
x_1(t)\\
x_2(t)
\end{array}\right)
-\hat p_t(x_1(t)+x_2(t))$.
The innovation process $W(t)$ is given by $W(t)=y(t)-\frac{1}{\sqrt{T_1}}\int_0^tx_1(s)+x_2(s)ds$.

\section{Conclusions}
In this paper, an approach to solving the problem of fault tolerant quantum filtering and fault detection for a class of laser-atom open quantum systems has been developed. A quantum-classical Bayesian inference method is considered to enable us to derive the fault tolerant quantum filter and fault detection equation in a convenient way. By describing the stochastic fault process as a finite-state jump Markov chain and using a reference probability approach, a set of coupled stochastic differential equations satisfied by the conditional system states and fault process estimate are derived. An application to two-level quantum systems under Poisson type faults is also presented. In the example, we have assumed that the measurement efficiency is 1. It is also straightforward to extend our result to the case with the measurement efficiency $\eta<1$.


\section*{Acknowledgement}
The authors are grateful to Dr. Hendra Nurdin, Prof. Matthew James and the anonymous reviewers for their constructive and valuable comments that have helped us greatly improve the presentation of this paper. The authors would also like to thank Dr. Mankei Tsang for pointing out three useful references.

\section*{Appendix}
\emph{Proof of Theorem 3.1.} Since the inverse mapping of a $*-$isomorphism is also a $*-$isomorphism, from Lemma 3.1 one can always construct a $*-$isomorphism $\iota_0$ mapping the set of measurable functions on $\Omega=\{1,...,n_r\}$ to a commutative $*-$algebra on $\mathscr{H}_0$. Applying a $*-$isomorphism $\bar{\iota}=\iota_0\otimes I$ to both sides of (\ref{gquantumAutoRevision2}) yields
\begin{equation}
\bar{\iota}(A_R)=\sum_{k=1}^{n_r}\nu(R_k)P_{R_k}\otimes U_{R_k}^{\dagger}AU_{R_k},\label{gquantumAutoRevision4}
\end{equation}
where $P_{R_k}=\iota_0(\textbf{1}_{R=R_k})$. Then $\bar{\iota}(A_R)$ is an operator on $\bar{\mathscr{H}}$. It can be verified that $P_{R_j}P_{R_k}=\iota_0(\textbf{1}_{R=R_j}\textbf{1}_{R=R_k})=\iota_0(\delta_{jk}\textbf{1}_{R=R_k})=\delta_{jk}P_{R_k}$, and $\sum_{k=1}^{n_r}P_{R_k}=\iota_0(\sum_{k=1}^{n_r}\textbf{1}_{R=R_k})=I$. Thus $\{P_{R_k}\}$ form a complete set of projection operators on $\mathscr{H}_0$. In addition, from Lemma 3.1, one can find a density operator $\bar \rho$ on $\mathscr{H}_0$ such that $\Tr(\bar \rho P_{R_k})=\mathbb{E}_{\mathcal{P}}(\textbf{1}_{R=R_k})$. Thus we have $\tilde{\mathbb{P}}(A_R)=\sum_{k=1}^{n_r}\nu(R_k)\mathbb{E}_{\mathcal{P}}(\textbf{1}_{R=R_k})\mathbb{P}(U_{R_k}^{\dagger}AU_{R_k})=\sum_{k=1}^{n_r}\nu(R_k)\Tr(\bar \rho P_{R_k})\Tr(\rho U_{R_k}^{\dagger}AU_{R_k})=\bar{\mathbb{P}}(\bar{\iota}(A_R))$.

Let $\bar{\mathscr{R}}=\{X|X=f(\bar{\iota}(A_R)), f: \mathbb{R}\to \mathbb{C}\}$ be a commutative $*-$algebra on $\bar{\mathscr{H}}$. Then from the above analysis we know the $*-$isomorphism $\bar{\iota}$ maps $\tilde{\mathscr{A}}$ onto $\bar{\mathscr{R}}$. The proof is thus completed. $\hspace{6.5cm}\qed$

\emph{Proof of Lemma 4.1.} Let $\tilde{\mathbb{Q}}^t$ be a normal state as $\tilde{\mathbb{Q}}^t(X)=\tilde{\mathbb{P}}(U^{\dagger}(t)XU(t))$. Let $K(t)$ be any element of $\mathscr{Y}_t$, then $K(t)=U^{\dagger}(t)K_o(t)U(t)$ for some $K_o(t)\in \mathscr{Q}_t$. Note the scalar valued function $\left<F(t),e_j\right>\in \mathscr{Q}_t'$ and $X\in \mathscr{Q}_t'$. We have
\begin{eqnarray}
&&\tilde{\mathbb{P}}(\tilde{\mathbb{P}}(\left<F(t),e_j\right>U^{\dagger}(t)XU(t)|\mathscr{Y}_t)K)\nonumber\\
&=&\tilde{\mathbb{P}}(\left<F(t),e_j\right>U^{\dagger}(t)XU(t)K(t))\nonumber\\
&=&\tilde{\mathbb{P}}(\left<F(t),e_j\right>U^{\dagger}(t)XK_o(t)U(t))\nonumber\\
&=&\tilde{\mathbb{Q}}^t(\left<F(t),e_j\right>XK_o(t))=\tilde{\mathbb{Q}}^t(\tilde{\mathbb{Q}}^t(\left<F(t),e_j\right>XK_o(t)|\mathscr{Q}_t))\nonumber\\
&=&\tilde{\mathbb{Q}}^t(\tilde{\mathbb{Q}}^t(\left<F(t),e_j\right>X|\mathscr{Q}_t)K_o(t))\nonumber\\
&=&\tilde{\mathbb{P}}(U^{\dagger}(t)\tilde{\mathbb{Q}}^t(\left<F(t),e_j\right>X|\mathscr{Q}_t)K_o(t)U(t))\nonumber\\
&=&\tilde{\mathbb{P}}(U^{\dagger}(t)\tilde{\mathbb{Q}}^t(\left<F(t),e_j\right>X|\mathscr{Q}_t)U(t)K(t)). \label{app1}
\end{eqnarray}
Letting $K(t)=(\tilde{\mathbb{P}}(\left<F(t),e_j\right>U^{\dagger}(t)XU(t)|\mathscr{Y}_t)\\-U^{\dagger}(t)\tilde{\mathbb{Q}}^t(\left<F(t),e_j\right>X|\mathscr{Q}_t)U(t))^{\dagger}$ yields
\begin{eqnarray}
&&\tilde{\mathbb{P}}(\left<F(t),e_j\right>U^{\dagger}(t)XU(t)|\mathscr{Y}_t)\nonumber\\
&=&U^{\dagger}(t)\tilde{\mathbb{Q}}^t(\left<F(t),e_j\right>X|\mathscr{Q}_t)U(t) \label{app2}
\end{eqnarray}
almost surely under $\tilde{\mathbb{P}}$

In addition, suppose the system is initialized at $\pi_0=\sum \limits_{k} p_k \left|\alpha_k\right>\left<\alpha_k\right|$ and we define a curve $\left|\psi_k(t)\right>=U(t)(\left|\alpha_k\right>\otimes \left|\upsilon \right>)$. Using the fact that $dB(t)\left|\upsilon \right>=0$,  one obtains (see Equation (6.13) in (\cite{Holevo1991}))
\begin{equation}
d\left|\psi_k(t)\right>=\{(-iH(F(t))-\frac{1}{2}L^{\dagger}L)dt+LdQ(t)\}\left|\psi_k(t)\right>. \label{app3}
\end{equation}
In other words, $U(t)(\left|\alpha_k\right>\otimes \left|\upsilon \right>)=V(t)(\left|\alpha_k\right>\otimes \left|\upsilon \right>)$ since $U(0)=V(0)=I$. After some mathematical manipulation, one obtains $\Tr(\rho_0 U^{\dagger}(t)XU(t))=\Tr(\rho_0 V^{\dagger}(t)XV(t))$ which leads to
\begin{equation}
\tilde{\mathbb{P}}(\left<F(t),e_j\right>U^{\dagger}(t)XU(t))=\tilde{\mathbb{P}}(\left<F(t),e_j\right>V^{\dagger}(t)XV(t)). \label{app4}
\end{equation}
Applying Theorem 3.2 by replacing $\Lambda$ with 1, $X$ with $\left<F(t),e_j\right>X\in \mathscr{Q}_t'$, $V$ with $V(t)$ and $\tilde{\mathscr{C}}$ with $\mathscr{Q}_t$ respectively yields
\begin{eqnarray}
\tilde{\mathbb{Q}}^t(\left<F(t),e_j\right>X|\mathscr{Q}_t)=\frac{\tilde{\mathbb{P}}(\left<F(t),e_j\right> V^{\dagger}(t)XV(t)|\mathscr{Q}_t)}{\tilde{\mathbb{P}}( V^{\dagger}(t)V(t)|\mathscr{Q}_t)}.  \label{app5}
\end{eqnarray}
Lemma 4.1 can be concluded by combining (\ref{app2}) and (\ref{app5}). $\qed$

\end{document}